\documentclass[preprints,article,accept,moreauthors,pdftex]{Definitions/mdpi}

\firstpage{1} 
\pubvolume{xx}
\issuenum{1}
\articlenumber{5}
\pubyear{2019}
\copyrightyear{2019}
\history{~}

\Title{Effects of turbulent environment on self-organized critical behavior: Isotropy vs Anisotropy}

\Author{N. V. Antonov $^{1}$, N. M. Gulitskiy $^{1}$, P. I. Kakin $^{1}$ and G. E. Kochnev $^{1}$}

\address[1]{%
$^{1}$ \quad Department of Physics,  
Saint-Petersburg State University, 7/9~Universitetskaya nab., St. Petersburg, 199034 Russia; n.antonov@spbu.ru, n.gulitskiy@spbu.ru, p.kakin@spbu.ru}

\abstract{We study a self-organized critical system under influence of turbulent motion of the environment. The system is described by the anisotropic continuous stochastic equation proposed by Hwa and Kardar [{\it Phys. Rev. Lett.} {\bf 62}: 1813 (1989)]. The motion of the environment is modelled by the isotropic Kazantsev--Kraichnan ``rapid-change'' ensemble for an incompressible fluid: it is Gaussian with vanishing correlation time and the pair correlation function of the form $\propto\delta(t-t’) / k^{d+\xi}$, where $k$ is the wave number and $\xi$ is an arbitrary exponent with the most realistic values $\xi = 4/3$ (Kolmogorov turbulence) and $\xi \to 2$ (Batchelor's limit). Using the field-theoretic renormalization group, we find infrared attractive fixed points of the renormalization group equation associated with universality classes, i.e., with regimes of critical behavior. The most realistic values of the spatial dimension $d=2$ and the exponent $\xi=4/3$ correspond to the universality class of pure turbulent advection where the nonlinearity of the Hwa--Kardar (HK) equation is irrelevant. Nevertheless, the universality class where both the (anisotropic) nonlinearity of the HK equation and the (isotropic) advecting velocity field are relevant also exists for some values of the parameters $\varepsilon=4-d$ and $\xi$. Depending on what terms (anisotropic, isotropic, or both) are relevant in specific universality class, different types of scaling behavior (ordinary one or generalized) are established.}


\begin{document}

\section{Introduction. \label{sec:Int}}

While equilibrium system can become critical only if a certain control
parameter is tuned to a precise critical value~\cite{Amit,Zinn,Book3}, systems with self-organized criticality (SOC)~\cite{BTW,BTW1,BTW2,Bak,Bak1,Bak2,Bak3} evolve towards critical state owing only to their intrinsic dynamics. The phenomenon of SOC attracted much attention~\cite{Col0,Col1,Col2,Col3} as it can be found in wide range of open non-equilibrium systems with dissipative transport. The examples include biological and neural systems~\cite{bio1,bio2,neu1,neu2,neu3,neu4,neu5,neu6}, online social networks~\cite{net1,net2,net3,net4,net5,net6}, and even agricultural systems~\cite{crop}. As data analysis grows more and more sophisticated, new instances of SOC are discovered. For example, the concept of SOC was applied to crisis behavior in autism spectrum disorders in~\cite{autism}.

The models of SOC are usually discrete. However, one can use continuous models instead to study critical scaling behavior (power law asymptotic behavior in long-time and large-distance range). Indeed, the continuous $\varphi^4$ model adequately describes universality classes (types of critical behavior) of several discrete models of equilibrium critical systems~\cite{Amit,Zinn,Book3}. Similarly, non-equilibrium universality class of kinetic roughening is successfully studied with the continuous Kardar--Parisi--Zhang \mbox{model~\cite{FNS,KPZ}}. There are continuous models for conserved directed-percolation processes which are related to the Manna universality class of SOC~\cite{SDP1,SDP2,SDP3}.

One of the continuous models of SOC was proposed by Hwa and Kardar (HK) in~\cite{HK,HK1}. It is a stochastic differential equation with anisotropy that models a ``running'' sandpile, i.e., a surface of a sandpile (flat on average) with a constant tilt. Influx of sand into the system causes avalanches and the surface becomes self-similar. The tilt provides the preferred direction for sand transport (and, thus, the  anisotropy).

Let us describe the HK model first. The scalar field $h(x)=h(t,{\bf x})$ stands for a deviation of the surface height from its average value ($t$ and ${\bf x}$ denote time-space coordinates). The preferred direction in $d$-dimensional space is defined by the unit constant vector $ {\bf n}$ that sets a decomposition for any vector~${\bf p}$: ${\bf p} = {\bf p}_{\perp} + {\bf n}\, p_{\parallel}$ where
$({\bf p}_{\perp} \cdot \, {\bf n}) =0$. The spatial derivative ${\bf \partial}=\partial/ \partial {x_{i}}$, $i=1,\dots, d$, is, thus, replaced with a $(d-1)$-dimensional derivative ${\bf \partial_{\perp}}=\partial/ \partial {x_{i}}$, $i=1,\dots, (d-1)$, and a one-dimensional derivative
$\partial_{\parallel} = ({\bf n} \cdot {\bf \partial})$.

The stochastic differential equation for the field $h(x)$ is
\begin{equation}
\partial_{t} h= \nu_{\perp 0}\, {\bf \partial}_{\perp}^{2} h + \nu_{\parallel 0}\,
\partial_{\parallel}^{2} h -
\partial_{\parallel} h^{2}/2 + f.
\label{eq1}
\end{equation}
Here $\partial_{t} = \partial/ \partial {t}$, ${\bf \partial}_{\perp}^{2}=({\bf \partial}_{\perp}\cdot{\bf \partial}_{\perp})$, $\nu_{\parallel 0}$ and
$\nu_{\perp 0}$ are diffusivity coefficients, and $f(x)$ is a Gaussian random noise with zero mean and the correlation function
\begin{equation}
\langle f(x)f(x') \rangle = D_{0}\,
\delta(t-t')\, \delta^{(d)}({\bf x}-{\bf x}'), \quad
\label{forceD}
\end{equation}
where $D_0$ is a positive amplitude.\footnote{
Traditionally, the nonlinear term $\partial_{\parallel} h^{2}/2$ has a coupling constant as a prefactor. Here, the fields and the parameters were scaled to make this factor equal to unity (the coupling constant, thus, appears in the amplitude of the correlation function for the random noise $f$).}

The HK model with a random field for a coupling constant was studied in~\cite{Tadic}. The HK model was also discussed in connection with erosion of landscapes in~\cite{Pastor1,Pastor2} and considered in connection with another type of random noise in~\cite{Vestnik,Stat,Serov,SerovEP}.

Turbulent motion of the environment and other external disturbances can drastically change critical scaling behavior; see, e.g., \cite{Satten,Satten1,Onuki,Onuki2,Beysens,Ruiz,Nelson,AHH,Alexa,AIK,AKM}. Moreover, simple estimation shows that the fully developed turbulence with the Reynolds number \mbox{Re$\gg1$} can be present in ordinary laboratory and atmospheric conditions.
Thus, we are interested in the effects of turbulent advection on a system described by the HK model. 
Some example of the physical systems of such kind may be a sand seabed under a troubled water. 

The advection by the velocity field ${\bf v}(x)$ is introduced by the ``minimal'' replacement \mbox{$\partial_{t} \to \nabla_{t} = \partial_{t} + ({\bf v}\cdot {\bf \partial})$} in Eq.~(\ref{eq1}). 
The new Lagrangian derivative $\nabla_{t}$ is Galilean covariant.
Let us model the velocity field ${\bf v}(x)$ by statistical ensemble with zero mean and the correlation function of the form:
\begin{equation}
\begin{aligned}
\langle v_{i} (t, {\bf x}) v_{j}(t',{\bf x}')\rangle =  \delta(t-t')
D_{ij}({\bf x}-{\bf x}'),
\\
D_{ij}({\bf r}) = B_{0}
\int_{k>m} \frac{d{\bf k}}{(2\pi)^{d}} \frac{1}{k^{d+\xi}}
 P_{ij}({\bf k})
\exp ({\rm i} {\bf k}\cdot{\bf r}).
\label{white}
\end{aligned}
\end{equation}
This is the celebrated Kazantsev--Kraichnan ``rapid-change'' ensemble~\cite{FGV}.
Here $P_{ij}({\bf k}) = \delta_{ij} - k_i k_j / k^2$ is the transverse
projector, $k\equiv |{\bf k}|$ is the wave number,
$B_{0}>0$ is a positive amplitude. 
The presence of the transverse projector $P_{ij}({\bf k})$ provides incompressibility of fluid, i.e., the property $({\bf  \partial}\cdot {\bf v}) = 0$. The $\delta$-function in $t$ ensures vanishing correlation time. 

The cutoff $k>m$ serves as an infrared (IR) regularization. From physics viewpoints, the parameter $m$ has the meaning of the inverse integral (outer) turbulence scale: $m\sim L^{-1}$. In laboratory or atmospheric turbulence $L$, the largest scale in the problem, can be estimated  as~$1\div100$~m; see, \mbox{e.g., \cite{Legacy,Ziad1,Ziad2}}. The precise form of the IR regularization is unimportant, because the critical dimensions
we are interested in here do not depend on its choice~\cite{Book3}.
Another type of regularization $k^2\to k^2+m^2$ is possible and was employed in earlier works on the Kraichnan model;
see, e.g., \cite{FGV} and references therein.
The sharp cutoff we use here is more convenient for calculational reasons.

The exponent $\xi$ usually spans the range of $[0,2]$ and can be viewed as a measure of the velocity ``roughness'' (H\"{o}lder exponent) with the limit $\xi\to2$ corresponding to a completely smooth field (Batchelor's limit). The most realistic ``Kolmogorov'' value is $\xi=4/3$. 

The ensemble can be ``derived'' from the stochastic Navier--Stokes (NS) equation as follows~\cite{Ant1,Ant2}.
Neglect of the nonlinearity and introduction of an effective viscosity coefficient reduces the NS equation to a linear one leading to the Ornstein--Uhlenbeck process with Gaussian statistics~\cite{Ito1,Ito2}. The correlation function~(\ref{white}) is then obtained as a certain limit of the model parameters.
Although the Kazantsev--Kraichnan ensemble is ``synthetic'' and rather simple,  it played a very important role in the study of intermittency and anomalous
scaling in fully developed fluid turbulence; see the review paper~\cite{FGV} and references therein.
It was also used to investigate the effect of turbulent advection on the critical scaling behavior in~\cite{AHH,Alexa,AIK,AKM}.

Since the model~(\ref{eq1}) is anisotropic, the first avenue to explore is to consider it with 
a velocity field that is also anisotropic. This was done in~\cite{AK1} where $d$-dimensional generalization of the Avellaneda and Majda ensemble~\cite{AM,AM1} was used. In addition to universality class of ordinary diffusion and universality class where only nonlinearity of the HK equation is relevant in the sense of Wilson, universality class of pure turbulent advection was established. No regime of critical behavior where both nonlinearity and advection are relevant was found (except for a very special choice of the model parameters).

In this paper, we choose the isotropic Kazantsev--Kraichnan ensemble instead. This isotropic ensemble is more natural from a physics viewpoint and can be interpreted as an intermediate step before considering coupling with the NS equation. Moreover, interplay between anisotropy and isotropy may provide  interesting and unexpected results. 

We use the field theoretic renormalization group (RG) approach in which the original stochastic problem~(\ref{eq1})~-- (\ref{white}) is replaced with an equivalent field theory. The regimes of IR asymptotic behavior 
(universality classes) are associated with IR attractive fixed points of 
the RG equations.  

The paper is organized as follows. In Sec.~\ref{sec:QFT1} we give a short summary of the RG analysis of the HK model performed in~\cite{HK,HK1}. We outline the analysis in terms of the field theoretic RG so that we can compare it with the analysis of the model with turbulent advection. This comparison allows us to observe subtle influence of isotropy of the velocity ensemble on symmetries, canonical dimensions, etc. We also discuss scaling behavior and present a derivation of the critical exponents. 

Secs.~\ref{sec:QFT2}~-- \ref{sec:scal} are devoted to the model with turbulent advection described by Eqs.~(\ref{eq1})~-- (\ref{white}).
In Sec.~\ref{sec:QFT2} we renormalize the constructed field theory for the model and calculate renormalization constants in the leading order of the double expansion in $\varepsilon=4-d$ and $\xi$. Sec.~\ref{sec:FP} is reserved for the analysis of RG functions and IR attractive fixed points. 
Sec.~\ref{sec:scal} is devoted to critical scaling and critical dimensions of the system in different scaling regimes. All obtained results are compared with the known special cases. Sec.~\ref{sect:Con} is reserved for conclusion. 

The main result is that the system displays four different regimes of critical behavior. Depending on the values of $\varepsilon$ and $\xi$, there are four possible regimes: the regime where critical properties of the system are defined 
by the mean field theory (ordinary diffusion), the regime where only the turbulent advection is relevant, the regime where only the nonlinearity of the HK equation is relevant, and, finally, the regime where both the advection and the nonlinearity are relevant simultaneously. 
In particular, the most realistic values of the spatial dimension $d=2$ and the exponent $\xi=4/3$
correspond to the regime of the pure turbulent advection where both the nonlinearity of the HK equation and its anisotropy become irrelevant as if
they are ``washed away'' by the turbulent flow. The scaling is ``isotropic'' in the sense that the spatial coordinates $x_{\parallel}$ and ${\bf x_{\bot}}$ are rescaled in an identical way as a single set ${\bf x}=\{x_{\parallel},{\bf x_{\bot}}\}$.

The scaling behavior that corresponds to the regime where only the nonlinearity of the HK equation is relevant (the limit of the pure HK model) is realized through
a kind of ``dimensional transmutation'': the ratio $u$ of the two diffusivity coefficients $\nu_\parallel$ and $\nu_\perp$ acquires in this limit a nontrivial canonical dimension. The regime where both the turbulent advecting field (the isotropic one) and the nonlinearity (the anisotropic one) are relevant appears to be the most intriguing. Here the ``ordinary'' scaling without rescaling of IR irrelevant parameters is prohibited but ``restricted'' scaling with rescaling of the times, coordinates, and mentioned above ratio $u$ is possible. 
This resembles modified types of scaling hypotheses (weak scaling in the spirit of Stell and generalized scaling in the spirit of Fisher) for systems which have more than one significantly different characteristic scales~\cite{Stell,Stell2,Stell3,Stell4}.

\section{RG analysis of the HK model without turbulent advection.\label{sec:QFT1}}

Let us first consider the HK model without turbulent advection. The RG analysis of the model was performed in~\cite{HK,HK1}.
Those authors applied the dynamic RG in the Wilsonian formulation.
Here we reproduce the results of their analysis in terms of the field theoretic RG for the ease of comparison with the RG analysis of the model with turbulence studied in this paper.

According to the general statement (see, e.g., Sec.~5.3 in the monograph~\cite{Book3} and references therein), the
stochastic problem~(\ref{eq1})~-- (\ref{forceD}) is equivalent\footnote{The equivalence means that the correlation and response functions of the problem~(\ref{eq1})~-- (\ref{forceD}) can be identified with various Green functions of the field theory with the action~(\ref{action_pure}). In other words, the correlation functions are represented by the functional averages over the initial field $h$ and the auxiliary (response) field $h'$ with the weight $\exp S(\Phi)$; for more details, see~\cite{Book3}.} 
to the field theory with the doubled set of fields $\Phi=\{h,h'\}$ and the action functional 
\begin{equation}
\begin{aligned}
    S(\Phi) = \frac{1}{2}h'D_0\,h' + h'(-\partial_th + \nu_{\parallel0}\,\partial^2_{\parallel}h + \nu_{\perp0}\,{\bf \partial}^2_{\perp}h - \frac{1}{2}\,\partial_{\parallel}h^2).
    \label{action_pure}
\end{aligned}    
\end{equation} 
Here and below we use symbolic notation in which integration over the arguments $x=\{t,{\bf x}\}$ is implied for all terms in expressions for action functionals, e.g.,
\begin{equation}
\frac{1}{2}\, h'D_0\,h' = \frac{1}{2}\,D_0\int dt\int d{\bf x}\, h'(t,{\bf x})h'(t,{\bf x}).
\end{equation} 

In general, dynamical models have two scales: the temporal scale $T$ and the spatial scale $L$. Thus, the canonical dimension of some quantity $F$ is described by two numbers: the momentum dimension $d^{k}_{F}$ and the frequency dimension $d^{\omega}_{F}$; see, e.g., Sec.~5.14 in~\cite{Book3}.
However, some special cases may require more accurate dimensional analysis. For example, in anisotropic models like~(\ref{eq1}) two independent spatial scales $L_{\parallel}$ and $L_{\perp}$, related to the preferred direction ${\bf n}$ and the 
orthogonal subspace, can be introduced; see, e.g., Sec.~1.17 in~\cite{Book3}. Thus, any quantity $F$ is described by three canonical dimensions, two of which ($d^{\parallel}_{F}$ and $d^{\perp}_{F}$) are related to the momentum scales:
\begin{equation}
    [F] \sim [T]^{-d^{\omega}_{F}}[L_{\parallel}]^{-d^{\parallel}_{F}}[L_{\perp}]^{-d^{\perp}_{F}}.
\end{equation}
By assuming that each term in the action~(\ref{action_pure}) is dimensionless and by using the obvious normalization conditions
\begin{equation}
d_{k_{\parallel}}^{{\parallel}} = -d_{x_{\parallel}}^{{\parallel}} = 1, d_{k_{\perp}}^{{\perp}} = -d_{x_{\perp}}^{{\perp}} = 1,  d_{k_{\parallel}}^{{\perp}} = -d_{x_{\parallel}}^{{\perp}} = 0, d^{\omega}_{k_{\parallel}} = d^{\omega}_{k_{\perp}}=0, d_{\omega}^{\omega}=-d_t^{\omega}=1
\label{normalization_canon}
\end{equation}
one can determine canonical dimensions of any quantity $F$.

The total canonical dimension $d_{F}$ is defined by the expression 
$d_{F} = d_{F}^{k} + 2d_{F}^{\omega}$, where $d_F^k=d_{F}^{\parallel}+d_{F}^{\perp}$ is the total momentum canonical dimension (the factor $2$ follows from the fact that $\partial_t \propto {\bf \partial}^2$ in the free theory). In the renormalization of dynamical models  $d_{F}$  plays the same role as the conventional momentum dimension is static models; see, e.g., \cite{Book3}.
Canonical dimensions of the fields and the parameters for the model~(\ref{action_pure}) are presented in Table~\ref{canon_dim_pure}; 
some of these parameters will be introduced later on.

\begin{table}[t]
\centering
\begin{tabular}{ | c | c | c | c | c | c | c | c | c | c |}
\hline
$F$ & $h'$ & $h$ & $D_0$ & $\nu_{\parallel0}$ &$ \nu_{\perp0}$& $u_0$&$g_0$&$g$&$\mu$,$m$,$\Lambda$      \\ \hline
$d_F^{\omega}$&$-1$&$1$&$3$&$1$&$1$&$0$&$0$&$0$&$0$\\ \hline
$d_F^{\parallel}$&$2$&$-1$&$-3$&$-2$&$0$&$-2$&$0$&$0$&$0$\\ \hline
$d_F^{\perp}$&$d-1$&$0$&$1-d$&$0$&$-2$&$2$&$\varepsilon$&$0$&$1$\\ \hline
$d_F^{k}$&$d+1$&$-1$&$-d-2$&$-2$&$-2$&$0$&$\varepsilon$&$0$&$1$\\ \hline
$d_F$&$d-1$&$1$&$4-d$&$0$&$0$&$0$&$\varepsilon$&$0$&$1$\\\hline
\end{tabular}
\caption{Canonical dimensions of the fields and the parameters in the theory~(\ref{action_pure}); $\varepsilon=4-d$.}
\label{canon_dim_pure}
\end{table}

Interaction vertex of the model~(\ref{action_pure}) corresponds to the nonlinear term $-h'\partial_{\parallel}h^2/2$. Traditionally, the vertex contains a coupling constant as a prefactor. 
However, if both the vertex and the correlation function~(\ref{forceD}) have independent
amplitudes, the model involves ``too many parameters'' and the canonical dimensions cannot be determined in a unique way. It is convenient, then, to rescale the fields and the parameters  so that the ``superfluos'' factor in the vertex is removed. As a result,  the coupling constant appears in the term $h'D_0\,h'/2$. More precisely it is defined by the relation
\begin{equation}
    D_0 = g_0\nu_{\perp0}^{3/2}\nu_{\parallel0}^{3/2}.
    \label{g_0_definition}
\end{equation}
From the canonical dimension analysis it follows that  $g_0\sim\Lambda^{\varepsilon}$, where $\Lambda$ is a certain characteristic ultraviolet (UV) momentum scale and $\varepsilon= 4-d$. Thus, the model is logarithmic (the coupling constant~$g_0$ becomes dimensionless) at $\varepsilon=0$, i.e., at $d=4$.
It is also convenient to introduce the quantity $u_0 = \nu_{\parallel0}/\nu_{\perp0}$. Here it is a dimensional parameter that possesses nontrivial momentum canonical dimensions $d_u^\parallel$ and $d_u^\perp$; see~Table~\ref{canon_dim_pure}.

An important role in the RG analysis of the HK model is played by its symmetries. Firstly, the action~(\ref{action_pure}) is invariant with respect to the reflection
\begin{equation}
x_{\parallel}\rightarrow -x_{\parallel}, \quad h'(t,\{x_{\parallel},{\bf x_{\perp}}\})\rightarrow -h'(t,\{-x_{\parallel},{\bf x_{\perp}}\}), \quad h(t,\{x_{\parallel},{\bf x_{\perp}}\})\rightarrow-h(t,\{-x_{\parallel},{\bf x_{\perp}}\}).
\label{sym2}
\end{equation}
Secondly, there is a tilt symmetry which can be viewed as the Galilean symmetry restricted to the subspace along the preferred direction ${\bf n}$: 
\begin{equation}
{\bf x}\rightarrow{\bf x} + Ut{\bf n},\quad h'(t,{\bf x})\rightarrow h'(t,{\bf x}+ Ut{\bf n}), \quad h(t,{\bf x})\rightarrow 
h(t,{\bf x}+ Ut{\bf n})-U, \quad U=\mbox{const}.
\label{sym1}
\end{equation}
The possible counterterms must satisfy these symmetries. What is more,   
any counterterm can involve the field $h'$ only in the form of a derivative $\partial_{\parallel}h'$
(or can be reduced to such form using the integration by parts).
Then the standard analysis of UV divergences based on the canonical dimensions (“power counting”)~\cite{Book3,Zinn} shows that there is only one such counterterm and it has the form $h'\partial_{\parallel}^{2}h$; it eliminates UV divergences from the 1-irreducible Green function 
$\langle h'h\rangle_{1-ir}$.

As a consequence, the model~(\ref{action_pure}) is multiplicatively renormalizable with only one nontrivial renormalization constant $Z_{\nu_{\parallel}} $\cite{HK,HK1}:
\begin{equation}
    S_R(\Phi) = \frac{1}{2}h'D\,h' + h'(-\partial_th + Z_{\nu_{\parallel}}\nu_{\parallel}\partial_{\parallel}^2h + \nu_{\perp}{\bf \partial}_{\perp}^2h - \frac{1}{2}\partial_{\parallel}h^2).
    \label{action_renorm_pure}
\end{equation}
Here the fields $h$, $h'$ and the diffusivity coefficient {$\nu_{\perp 0}=\nu_{\perp}$} 
are not renormalized ($Z_h=Z_{h'}=Z_{\nu_{\perp}}=1$)
while the bare parameters $\nu_{\parallel0}$, $D_0$, and $g_0$ are replaced with their renormalized counterparts:
\begin{equation}
\nu_{\parallel0} = \nu_{\parallel} Z_{\nu_{\parallel}} , \quad  
D_0 =D= g\mu^{\varepsilon}\nu_{\parallel}^{{3}/{2}}\nu_{\perp}^{{3}/{2}}, \quad
g_0= g\mu^{\varepsilon} Z_g,
\label{REN1}
\end{equation}
where the reference mass $\mu$ is an additional arbitrary parameter of the renormalized theory.

The canonical scale invariance of a certain renormalized Green function $G^R=\langle \Phi \dots \Phi \rangle$ is expressed by three differential equations
\begin{equation}
\begin{aligned}
    \left(\sum_i d_i^{\omega} {\cal D}_i - d_{G}^{\omega} \right)G^R &= 0, \\
    \left(\sum_i d_i^{\perp} {\cal D}_i - d_{G}^{\perp} \right)G^R &= 0,\\
    \left(\sum_i d_i^{\parallel} {\cal D}_i - d_{G}^{\parallel} \right)G^R &= 0.
\label{ScInvX}
\end{aligned}
\end{equation}
Here $i$ is the full set of all the arguments of $G^R$, namely, $\omega$, $k_{\perp}$, $k_{\parallel}$, $\mu$, $\nu_{\perp}$, and $\nu_{\parallel}$ (or equivalently $u$).
The canonical dimensions $d_G^{\perp}$, $d_G^{\parallel}$, and 
$d_G^\omega$ of the Green function $G^R$ 
in the coordinate representation
are given by simple sums of the corresponding dimensions of the fields entering into $G^R$;
the operator ${\cal D}_a$ is defined as ${\cal D}_a=a\partial_a$ for any variable $a$.

The basic differential RG equation has the form 
\begin{equation}
{\left(
{\cal D}_{\mu} + \beta_g\partial_g - \gamma_{\nu_\parallel}{\cal D}_{\nu_\parallel}
- \gamma_{\nu_\perp}{\cal D}_{\nu_\perp} - \gamma_G \right) G^{R} = 0.}
\label{RGeqX}
\end{equation}
In general, the anomalous dimensions $\gamma$ and $\beta$ functions 
(generally referred to as RG functions) are defined as 
\begin{equation}
    \gamma_{F} = \widetilde{\cal D}_{\mu}\ln Z_{F}, \quad  \beta_g = \widetilde{\cal D}_{\mu}g
    \label{anomal}
\end{equation}
for any quantity $F$ and any coupling constant $g$. Here $Z_{F}$ is the renormalization constant of $F$ and $\widetilde{\cal D}_{\mu}$ is the differential operator ${\cal D}_{\mu}=\mu\partial_{\mu}$ taken at fixed bare parameters. Since in the present model the fields $h$ and $h'$ and the parameter $\nu_\perp$ are not renormalized, one has $\gamma_{h}=\gamma_{h'}=\gamma_{\nu_{\perp}}=0$.

Possible types of asymptotic scaling regimes are determined by fixed points of the RG equations.  The coordinates of the latter are given by the zeroes of the $\beta$ functions, $\beta(g^*)=0$.
A fixed point is IR attractive (i.e., it corresponds to long-time, large-scale asymptotic behavior) if the real parts of all the eigenvalues $\lambda_i$ of the matrix $\Omega_{ij}=\partial_{g_i}\beta_{g_j}$ are positive, where $g =\{g_i\}$ denotes the full set of the coupling constants.

The substitution $g\to g^*$ and, hence, $\gamma_F \to\gamma_F ^*$ brings the RG equation to an equation with constant coefficients of the same type as Eqs.~\eqref{ScInvX}: 
\begin{equation}
\left(
{\cal D}_{\mu}  - \gamma^*_{\nu_\parallel}{\cal D}_{\nu_\parallel}
- \gamma^*_{\nu_\perp}{\cal D}_{\nu_\perp} - \gamma_G^* \right) G^{R} = 0.
\label{RGeqXXX}
\end{equation}
Each of the three equations~\eqref{ScInvX} and the equation~\eqref{RGeqXXX} describe a certain independent scaling behavior of the function $G^R$ in which some of its variables are rescaled and some are kept fixed. A parameter is rescaled if the corresponding derivative enters the differential operator; otherwise the parameter is fixed (for more details see, e.g., Sec.~1.2 in~\cite{Book3}).

We are interested in the critical scaling behavior where the frequencies and momenta (or, equivalently, times and coordinates) are rescaled, while the IR irrelevant parameters (namely, $\mu$, $\nu_{\perp}$
and $\nu_{\parallel}$) are kept fixed.\footnote{This corresponds to a real experimental setup, 
where for any specific material both diffusivity coefficients and $\mu$ (which is related to the characteristic intermolecular distance or some other microscopic length scale) are fixed.}
Thus, we combine all those equations to exclude 
the derivatives with respect to all the IR irrelevant parameters\footnote{As we will see in Sec.~\ref{sec:scal}, such an exclusion is not always possible: it requires some balance between the numbers of IR relevant and IR irrelevant parameters and the number of independent scaling equations.}
and arrive at the critical scaling equation for a given fixed point:
\begin{equation}
    \left({\cal D}_{k_\perp} +{\cal D}_{k_\parallel} \Delta_\parallel+ \Delta_\omega {\cal D}_\omega
        - \Delta_G  \right)G^{R} = 0
    \label{HK-scalingXZ}
\end{equation}
with $\Delta_{\parallel} = 1 + \gamma_{\nu_{\parallel}}^{*}/2$ and $\Delta_\omega=2-\gamma_{\nu_\perp}^*$ ($\Delta_\perp=1$ is the normalization condition). The critical dimension for the function $G^R$ is
\begin{equation}
    \Delta_G = d_{G}^{\perp} + d_{G}^{\parallel}\Delta_{\parallel} +\Delta_{\omega} d^{\omega}_G + \gamma_{G}^{*}, 
    \label{Dimens}
\end{equation}
where $d_{G}^{\perp}$, $d_{G}^{\parallel}$ and $d_{G}^{\omega}$ are the canonical dimensions of the function $G^R$ and $\gamma_{G}^{*}$ is its anomalous dimension
at the fixed point.

The critical dimensions characterize universality classes of the critical behavior and appear as exponents in the power law for critical behavior of the Green functions. In particular, for the pair correlation function, the solution of~\eqref{HK-scalingXZ} has the form
\begin{equation}
\langle h(t,{\bf x})\, h(0,{\bf 0}) \rangle \simeq
r_{\perp}^{-2\Delta_{h}}\, {\cal F} \left(t/ r_{\perp}^{\Delta_{\omega}},
r_{\parallel}/ r_{\perp}^{\Delta_{\parallel}} \right),
\label{dimS}
\end{equation}
where $r=|{\bf x}|$ and ${\cal F}$ is a certain scaling function; 
the dependence on the IR irrelevant parameters is omitted.

The notations $z=\Delta_{\omega}/\Delta_{\parallel}$, $\zeta=1/\Delta_{\parallel}$, and $\chi = -\Delta_{h}/\Delta_{\parallel}$ are used in~\cite{HK,HK1}. We recall that in the HK model~\eqref{action_pure} $\gamma_{G}^{*}=\gamma_{\nu_\perp}^*=0$, which in particular implies $\Delta_\omega=2$.

The model has two fixed points. The Gaussian (free) fixed point with $g^{*}=0$ and $\gamma_{\nu_{\parallel}}^{*}=0$ is IR attractive for $\varepsilon<0$; the corresponding
critical dimensions read
\begin{equation}
\Delta_h = 1, \quad \Delta_{h'} = 3 - \varepsilon, \quad \Delta_{\omega}=2, \quad
\Delta_{\parallel} = 1.
\end{equation}
The nontrivial fixed point with $g^*=32\varepsilon/9+O(\varepsilon^2)$ is IR attractive for $\varepsilon>0$. It corresponds to the regime of critical behavior where the nonlinearity of the HK model is relevant in the sense of Wilson.
Owing to the second relation in~(\ref{REN1}), the value $\gamma_{\nu_{\parallel}}^{*}=2\varepsilon/3$ is found exactly. Then the relations~(\ref{Dimens}) give
\begin{equation}
        \Delta_h = 1-\varepsilon/3, \quad \Delta_{h'} = 3 - \varepsilon/3, \quad 
        \Delta_{\omega}=2, \quad   \Delta_{\parallel} = 1+ \varepsilon/3.
        \label{HKcrit}
\end{equation}
All the above expressions are perturbatively exact, that is, they have no higher-order corrections in $\varepsilon$.

\section{Renormalization of the model with turbulent advection.\label{sec:QFT2}}

Inclusion of turbulent advection in the HK model (that is achieved by replacing $\partial_t$ with $\nabla_t$, see~Sec.~\ref{sec:Int}) adds two new terms and a new field ${\bf v}(x)$ into the action~\eqref{action_pure}: 
\begin{equation}
\begin{aligned}
    S(\Phi) = \frac{1}{2}h'D_0\,h' + h'(-\nabla_th + \nu_{\parallel0}\partial^2_{\parallel}h + \nu_{\perp0}\partial^2_{\perp}h - \frac{1}{2}\partial_{\parallel}h^2) + S_{{\bf v}}.
    \label{action_ve}
\end{aligned}    
\end{equation} 
\begin{equation}
    S_{{\bf v}} = -\frac{1}{2}\int dt \int d{\bf x} \int d{\bf x}' v_i(t,{\bf x})D^{-1}_{ij}({\bf x}-{\bf x}')v_j(t,{\bf x}').
\label{action_Sv}
\end{equation}
Here $D^{-1}_{ij}({\bf x} - {\bf x'})$ is the kernel of the inverse operator $D^{-1}_{ij}$ for the integral operator $D_{ij}$ from~(\ref{white}) as $S_{{\bf v}}$ is equivalent to Gaussian averaging over the field ${\bf v}$ with the correlation function~(\ref{white}).

In the frequency-momentum $(\omega-{\bf k})$ representation, the bare propagators for the theory~(\ref{action_ve}) have the following forms:
\begin{equation}
\begin{aligned}
    \langle hh' \rangle_0 &= \langle h'h \rangle_0^{*} = \frac{1}{-i\omega + \epsilon({\bf k})}, 
    \\
    \langle h'h' \rangle_0 &= 0, \quad
    \langle hh \rangle_0 = \frac{D_0}{\omega^2+\epsilon^2({\bf k})},
    \\
    \langle v_iv_j \rangle_0 &= B_0 \frac{P_{ij}({\bf k})}{k^{d+\xi}},
\label{propagators}
\end{aligned}
\end{equation}
where $\epsilon({\bf k}) = \nu_{\parallel0}{ k}_{\parallel}^2 + \nu_{\perp0}{\bf k}_{\perp}^2$.
Two vertices $-h'\partial_\parallel h^2/2$ and $-h'({\bf v}\cdot {\bf  \partial})h$ correspond to the vertex factors~$ik_\parallel^{h'}$ and $i{\bf k}^{h'}$, respectively; ${\bf k}^{h'}$ denotes momentum of the field $h'$.

Since the velocity ensemble~(\ref{white}) is isotropic, it is not possible to define two independent spatial scales in this model. Thus, 
\begin{equation}
    [F] \sim [T]^{-d^{\omega}_{F}}[L]^{-d^{k}_{F}}
\end{equation}
and $d_{F} = d_{F}^{k} + 2d_{F}^{\omega}$.
Canonical dimensions of the fields and parameters of the model~\eqref{action_ve} are presented in the Table~\ref{canon_dim_ve}.
We note that for the quantities that 
are present in the pure HK model with the action
(\ref{action_pure}) these dimensions coincide with their analogs in Table~\ref{canon_dim_pure}.

In contrast to the HK model itself (see Sec.~\ref{sec:QFT1})
and its modifications with anisotropic velocity
ensembles~\cite{AK1,Stat,Serov,SerovEP} (where the diffusivity coefficients $\nu_{\parallel0}$ and $\nu_{\perp0}$ have different momentum dimensions $d^\parallel$ and $d^\perp$), only total momentum dimension $d^k$ can be defined for the model~\eqref{action_ve}. 
We can see from Table~\ref{canon_dim_ve} that $d^{\omega}_{\nu_\parallel}=d^{\omega}_{\nu_\perp}$ and
$d^k_{\nu_\parallel}=d^k_{\nu_\perp}$.
As a result,
their ratio $u_0=\nu_{\parallel0}/\nu_{\perp0}$ appears to be completely dimensionless, that is, dimensionless with respect to the frequency and momentum dimensions separately.\footnote{That was not the case in the pure HK model, because the ratio $u_0$ was not dimensionless with respect to two possible momentum dimensions separately.} 
Therefore, according to the general rules, $u_0$ should be treated as an additional coupling constant.  
The third coupling constant in the model is $x_0\sim\Lambda^{\xi}$ related\footnote{
Since $\nu_{\parallel0}$ and $\nu_{\perp0}$ have the same dimensions, it is not possible at this point to determine which one (or both) should be used in the definition~\eqref{x_0_definition}.
This situation differs from Eq.~\eqref{g_0_definition} where $\nu_{\parallel0}$ and $\nu_{\perp0}$ have different canonical dimensions and, therefore, the exponents $3/2$ are strictly defined.
For simplicity, let us define $D_0$ the same way we did in Sec.~\ref{sec:QFT1} and define $B_0$ using Eq.~\eqref{x_0_definition}. The results do not depend on the specific realization of this arbitrariness.}
 to the amplitude $B_0$ of the correlation function~(\ref{white}):
\begin{equation}
    B_0 = x_0\nu_{\perp0}.
    \label{x_0_definition}
\end{equation}
From the Table~\ref{canon_dim_ve} it follows that the theory is logarithmic when $\varepsilon=0$ and $\xi=0$. Thus, the exponent $\xi$ which enters the correlation function~(\ref{white}) is a small parameter that should be considered alongside $\varepsilon$ in perturbation theory. 

\begin{table}[t]
\centering
\begin{tabular}{ | c | c | c | c | c | c | c | c | c | c | c | c | c |}
\hline
$F$ & $h'$ & $h$ & $D_0$ & $\nu_{\parallel0}$ &$ \nu_{\perp0}$&$g_0$&$v$&$B_0$&$x_0$&$u_0$,$u$&$g$,$x$&$\mu$, $m$, $\Lambda$ \\ \hline
$d_F^{\omega}$&$-1$&$1$&$3$&$1$&$1$&$0$&$1$&$0$&$0$&$0$&$0$&$0$\\ \hline
$d_F^{k}$&$d+1$&$-1$&$-2-d$&$-2$&$-2$&$\varepsilon$&$-1$&$\xi-2$&$\xi$&$0$&$0$&$1$\\ \hline
$d_F$&$d-1$&$1$&$4-d$&$0$&$0$&$\varepsilon$&$1$&$\xi$&
$\xi$&$0$&$0$&$1$\\\hline
\end{tabular}
\caption{Canonical dimensions of the fields and the parameters in the theory~(\ref{action_ve}); $\varepsilon=4-d$.}
\label{canon_dim_ve}
\end{table}

The model~(\ref{action_ve}) satisfies both symmetries of the model~(\ref{action_pure}) without advection granted that the symmetries are augmented to include the rules for the transformation of the field ${\bf v}$. The reflection symmetry~(\ref{sym2}) requires ${\bf v}(t,\{x_{\parallel},{\bf x_{\perp}}\}) \rightarrow -{\bf v}(t,\{-x_{\parallel},{\bf x_{\perp}}\})$ while the tilt symmetry~(\ref{sym1}) should simply be augmented by 
${\bf v}(t,{\bf x})\rightarrow {\bf v}(t,{\bf x}+ Ut{\bf n})$.

The action~(\ref{action_ve}) is also invariant with respect to another realization of the Galilean symmetry:
\begin{equation}
\begin{aligned}
&{\bf x}\rightarrow{\bf x} + Ut{\bf n},\quad h'(t,{\bf x})\rightarrow h'(t,{\bf x}+ Ut{\bf n}), \quad h(t,{\bf x})\rightarrow h(t,{\bf x}+ Ut{\bf n}),\\
&{\bf v}(t,{\bf x})\rightarrow {\bf v}(t,{\bf x}+ Ut{\bf n}) - U{\bf n}, \quad U=\mbox{const}.  
\end{aligned}
\end{equation}
Unlike the previous symmetries, this one is not strict: the action~(\ref{action_ve}) is invariant up to terms linear in the field ${\bf v}$ that stem from $S_{\bf v}$. Nevertheless, the counterterms must still satisfy the symmetry exactly. Indeed, the linear terms in the action variation could produce only disconnected and 1-reducible diagrams that do not give rise to new divergences; for more details see~\cite{Book3} (statement~3, Sec.~3.13) and~\cite{AntVestnik,AntVestnik2,AAKim}.

As before, the analysis of UV divergences and the discussed above symmetry considerations show that one has to introduce only the counterterms that eliminate the UV divergences from the 1-irreducible Green function $\langle h'h\rangle_{1-ir}$. However, the new vertex $-h'({\bf v}\cdot {\bf  \partial})h$ involves the full \mbox{derivative ${\bf  \partial}$}, so that the two counterterms $h'\partial_{\parallel}^{2}h$ and $h'\partial_{\perp}^{2}h$ are needed.

Thus, the renormalized action contains two nontrivial renormalization constants $Z_{\nu_{\parallel}}$ and $Z_{\nu_{\perp}}$:
\begin{equation}
    S_R(\Phi) = \frac{1}{2}h'D\,h' + h'(-\nabla_th + Z_{\nu_{\parallel}}\nu_{\parallel}\partial_{\parallel}^2h + Z_{\nu_{\perp}}\nu_{\perp}\partial_{\perp}^2h - \frac{1}{2}\partial_{\parallel}h^2) + S_{{\bf v}}.
    \label{action_renorm} 
\end{equation}
The fields $h$, $h'$, and ${\bf v}$ are not renormalized unlike the coupling constants $g_0$, $x_0$, $u_0$, and the diffusivity coefficients $\nu_{\parallel0}$, $\nu_{\perp0}$ which are related to their renormalized counterparts as follows:
\begin{equation}
\begin{aligned}
    \nu_{\parallel0} = \nu_{\parallel} Z_{\nu_{\parallel}} , \quad &\nu_{\perp0} = \nu_{\perp} Z_{\nu_{\perp}}, \\
    g_0=Z_g g\mu^{\varepsilon},\quad x_0=&Z_x x\mu^{\xi},\quad  u_0 = Z_u u.
    \label{renorm_param}
\end{aligned}    
\end{equation}
Here $Z_g$, $Z_x$, and $Z_u$ are renormalization constants for the coupling constants. The renormalized amplitudes  $D=D_0$ and $B=B_0$ are expressed in renormalized parameters as:
\begin{equation}
 D=\nu_{\parallel}^{3/2}\nu_{\perp}^{3/2}g\mu^{\varepsilon}, \quad B=x\nu_{\perp}\mu^{\xi}.
\label{D,B}
\end{equation}
The renormalization constants are related to each other as follows:
\begin{equation}
Z_g = Z_{\nu_{\parallel}}^{-3/2}Z_{\nu_{\perp}}^{-3/2}, \quad Z_x = Z_{\nu_{\perp}}^{-1}, \quad Z_u = Z_{\nu_{\parallel}}Z_{\nu_{\perp}}^{-1}.
\label{ZZZ}
\end{equation}
We omit the one-loop calculations of the renormalization constants for brevity; see., e.g., \cite{AKL} for similar calculations. 
In the one-loop approximation and minimal subtraction scheme (MS scheme) the renormalization constants $Z_{\nu_\parallel}$ and $Z_{\nu_\perp}$ read
\begin{equation}
    \begin{aligned}
    Z_{\nu_{\parallel}} &= 1 - \frac{3}{8u}\frac{x}{\xi}-\frac{3}{16}\frac{g}{\varepsilon}, \\
    Z_{\nu_{\perp}} &= 1 - \frac{3}{8}\frac{x}{\xi},
    \end{aligned}
    \label{zol}
\end{equation}
with the higher-order corrections in $g$ and $x$.\footnote{
To simplify the notation, here and below we redefine the coupling constants: $g\to gS_d/(2\pi)^d$ and $x\to x S_d/(2\pi)^d$, where $S_d=2\pi^d/\Gamma(d/2)$ is the unit sphere area in the $d$-dimensional space.} 

The anomalous dimensions (see definition~(\ref{anomal})), in their turn, have the following one-loop expressions:
\begin{equation}
    \gamma_{\nu_{\parallel}} = \frac{3}{8}\frac{x}{u} + \frac{3}{16}g, \quad \gamma_{\nu_{\perp}} = \frac{3}{8}x.
    \label{gamma1_gamma2}
\end{equation}
The anomalous dimensions for the coupling constants can be found using the relations~(\ref{ZZZ}) between renormalization constants:
\begin{equation}
    \gamma_{g} = - \frac{3}{2}\left(\gamma_{\nu_{\parallel}} + \gamma_{\nu_{\perp}}\right), \quad \gamma_x = -\gamma_{\nu_{\perp}}, \quad \gamma_u = \gamma_{\nu_{\parallel}} - \gamma_{\nu_{\perp}}.
    \label{an-2}
\end{equation}
Similarly to the case of the HK model without turbulent advection (see Sec.~\ref{sec:QFT1}), the anomalous dimensions for the fields $h$, $h'$, and ${\bf v}$ vanish due to the absence of their renormalization: \mbox{$\gamma_h=\gamma_{h'}=\gamma_v=0$}. 

From definitions~\eqref{anomal} and expressions~\eqref{an-2} it follows that the $\beta$ functions for the three coupling constants read
\begin{equation}
\begin{aligned}
\beta_g &= g\left(-\varepsilon +\frac{3}{2}\gamma_{\nu_{\parallel}} + \frac{3}{2}\gamma_{\nu_{\perp}}\right),\\
\beta_x &=x\left(-\xi+\gamma_{\nu_{\perp}}\right),\\
\beta_u &= u\left(-\gamma_{\nu_{\parallel}} + \gamma_{\nu_{\perp}}\right).
\end{aligned}
\label{beta}
\end{equation}

\section{Fixed points of the model with turbulent advection.\label{sec:FP}}

From explicit results for the anomalous dimensions~(\ref{gamma1_gamma2}), it follows that the  one-loop expressions for the $\beta$ functions have the form
\begin{equation}
\begin{aligned}
\beta_g &= g\left(-\varepsilon +\frac{9}{32}g+\frac{9}{16}\frac{x}{u}+\frac{9}{16}x\right),\\
\beta_x &=x\left(-\xi+\frac{3}{8}x\right),\\
\beta_u &= u\left(-\frac{3}{16}g-\frac{3}{8}\frac{x}{u}+\frac{3}{8}x\right).
\end{aligned}
\label{beta-expr1}
\end{equation}
The IR scaling behavior of the model is determined by the IR attractive fixed points of the RG equations, given by 
the set of equations $\beta_g=\beta_x=\beta_u=0$. Their analysis
reveals two possible IR attractive fixed points: the Gaussian point FP1 with the coordinates $g^*=0$, $x^*=0$, and arbitrary
$u^*$, and the fixed point FP2 with the coordinates $g^*=0$, $x^*=8\xi/3$, $u^*=1$ which corresponds to the regime of simple turbulent advection
(the HK nonlinearity is irrelevant in the sense of Wilson). 
It resembles the result of Onuki and Kawasaki~\cite{Onuki2}, where it was shown that the effects of fluid motion can destroy nontrivial critical behavior and turn it to the mean-field one.

The point FP1 is IR attractive for $\varepsilon<0$, $\xi<0$, whilst
the point FP2 is IR attractive for $\xi>\varepsilon/3$, $\xi>0$. This is the full set of fixed points with finite and non-zero value of $u^*$; there are no fixed points with $g^*\neq 0$ in the set, i.e., the HK universality class is not realized for such values of $u$. 

However,  the system~(\ref{beta-expr1}) may lose possible solutions with $u^*=0$ and $1/u^*=0$. 
In order to explore those exceptional values, we have to pass to new variables: $w=x/u$ (to study the case 
$u^*=0$) and to $\alpha=1/u$ (to study the case 
$u^*\to\infty$).\footnote{Our analysis is based on the perturbative expansion in 
$g$ and $x$. Thus, we do not try to consider possible 
fixed points with $1/g^*=0$ or $1/x^*=0$. In this connection we note that $1/g^*=0$ is not a fixed point for the $\phi^4$ model, where the large-$g$ behavior of the $\beta$ function is known~\cite{Kazakov}.}
For the first case we obtain
\begin{equation}
\beta_w=\widetilde{D}_\mu w=w\left(-\xi +\frac{3}{16}g+\frac{3}{8}w\right).
\end{equation}
Then the equations $\beta_g=\beta_w=\beta_u=0$ admit the solution  $u^*=0$ 
and give two additional fixed points: $g^* = 0$, $w^* = 8\xi/3$, $u^*=0$ and the point $g^* = 32\varepsilon/9$, $w^* = 0$, $u^*=0$, 
which corresponds to the HK model with $\varepsilon<0$. Nevertheless, both those points are saddle-type ones and cannot be IR attractive.

In the second case, we have
\begin{equation}
\beta_\alpha=\widetilde{\cal D}_\mu \alpha=\alpha\left(-\frac{3}{8}x+\frac{3}{8}x\alpha+\frac{3}{16}g\right).
\label{beta-a}
\end{equation}
Then the  system $\beta_g=\beta_x=\beta_\alpha=0$ leads to two more fixed points:
the point FP3 with the coordinates $g^*=32\varepsilon/9$, $x^*=0$, $\alpha^*=0$ and the point FP4 with  $g^*=32\varepsilon/9-16\xi/3$, $x^*=8\xi/3$, $\alpha^*=0$. From the physics point of view, the point FP3 corresponds to the regime of critical behavior where only the nonlinearity of the HK equation is relevant, and the point FP4 corresponds to the regime where both the nonlinearity
and the turbulent advection are relevant. The point FP3 is IR attractive if $\varepsilon>0$, $\xi<0$, and the point FP4 is IR attractive if $\xi<\varepsilon/3$, $\xi>0$. All other fixed points with $\alpha^*=0$ are not IR attractive at any values of $\xi$ and $\varepsilon$.
\begin{figure}[t]
    \centering
    \includegraphics[scale=0.6]{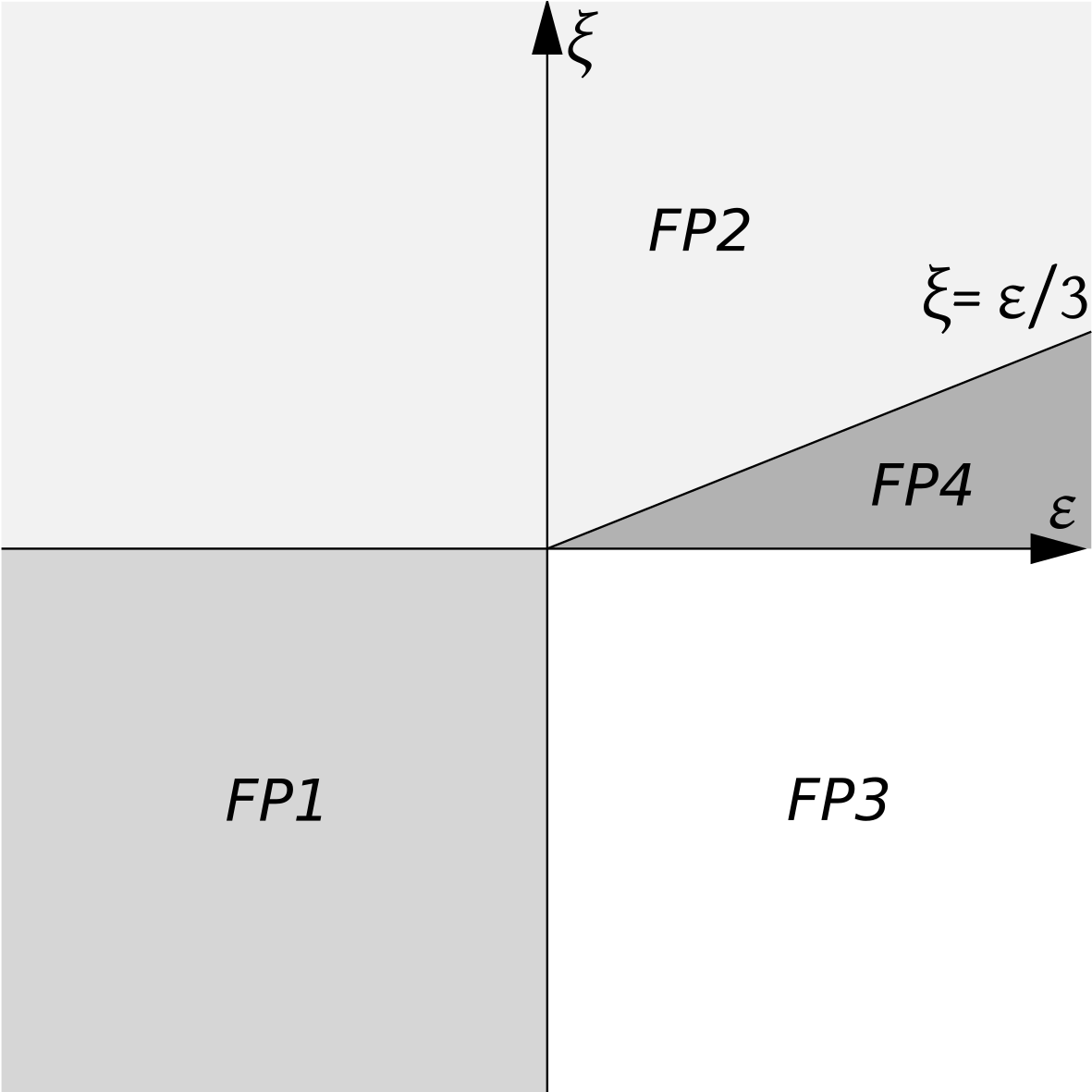}
    \caption{Regions of stability of the fixed points in the model~(\ref{action_ve}).
     Different areas correspond to the values of the parameters for which points FP1~-- FP4 are IR attractive. 
     The region FP1 corresponds to the mean field theory while the regions FP2, FP3, and FP4 correspond to the regimes where only the turbulent advection is relevant, only the nonlinearity of the HK equation is relevant, both the advection and the nonlinearity are relevant, respectively.} 
     \label{fig:diag_areas}
\end{figure}

The general stability pattern of the fixed points in the plane $(\varepsilon,\xi)$ is shown in Fig.~\ref{fig:diag_areas}. The straight line $\xi=\varepsilon/3$ denotes the border between the stability regions (areas where the
points are IR attractive) of the points FP2 and FP4. The stability regions of all points have neither gaps between
them nor overlaps. 
The Kolmogorov values $\xi=4/3$ and $\varepsilon=2$ (i.e., $d=2$) lie in the stability region of the point~FP2.

Existence of the IR attractive solutions of the RG equations leads to existence of the scaling behavior of the correlation functions. The possible scaling behavior and corresponding critical dimensions are discussed in the detail in the following Section.

\section{Scaling regimes and critical dimensions in the model with turbulent advection.\label{sec:scal}}

The general difference between the model under consideration and the original HK model~\cite{HK,HK1} (or the other models with anisotropic turbulent advection~\cite{AK1,Stat,Serov,SerovEP}) is that the former allows to introduce two different momentum dimensions while the latter does not. 
This means, that fully anisotropic models contain a system of three equations~(\ref{ScInvX}) related to the canonical scale invariance. At the same time, the model~\eqref{action_ve} contains only the following two equations:
\begin{equation}
    \begin{aligned}
    \left(\sum_i d_i^{\omega} {\cal D}_i - d_{G}^{\omega} \right)G^R &= 0, \\
    \left(\sum_i d_i^{k} {\cal D}_i - d_{G}^{k} \right)G^R &= 0. \\
    \end{aligned}
\label{ScInv}
\end{equation}

The differential RG equation for the theory~\eqref{action_ve} reads 
\begin{equation}
\left(
{\cal D}_{\mu} + \beta_g\partial_g + \beta_x\partial_x+ \beta_u\partial_u - \gamma_{\nu_\perp}{\cal D}_{\nu_\perp} -\gamma_G\right) G^{R} = 0.
\label{RGeqXX}
\end{equation}
When combined with~\eqref{ScInv} and taken at a fixed point, it yields
\begin{equation}
    \left({\cal D}_{k_\parallel}+{\cal D}_{k_\perp} +\Delta_\omega{\cal D}_\omega-d^k_G -\Delta_\omega d^\omega_G-\gamma^*_G\right)G^{R} = 0,
    \label{G-scaling}
\end{equation}
where $\Delta_\omega=2-\gamma_{\nu_\perp}^*$.
This is the equation of critical scaling and, as it should be, it does not contain derivatives over $\mu$ and $\nu_\perp$
(see Sec.~\ref{sec:QFT1} for discussion). 
The critical dimension $\Delta_F$ of a quantity $F$, therefore, reads
\begin{equation}
    \Delta_F=d^k_F +\Delta_\omega d^\omega_F+\gamma^*_F.
    \label{crete}
\end{equation}
For the fixed points FP2 (which incorporates the Kolmogorov value) critical dimensions are 
\begin{equation}
    \Delta_h=1-\xi, \quad \Delta_v=1-\xi, \quad \Delta_{h'}=3-\varepsilon+\xi, \quad \Delta_{\omega}=2-\xi.
\end{equation}
These expressions are perturbatively exact owning to the fact that $\gamma_{\nu_\perp}^*$ is known exactly while $\gamma_{h}=\gamma_{h'}=\gamma_{v}=0$. 
For similar reasons, the same feature is true for all the other fixed points of the model~(\ref{action_ve}).
As it should be, the value of the critical dimension $\Delta_v$ at this fixed point (where only the turbulent advection is relevant) coincides with the ones obtained for similar regimes in other problems where the same velocity ensemble was used; see, e.g., Eq.~(6.4) in~\cite{TMF64} and Eqs.~(2.18)~-- (2.20) in~\cite{AAV98}.

The fixed point FP1 is Gaussian, therefore, all the critical dimensions for this regime coincide with the canonical ones. 

The fixed points FP3 and FP4 appear to be more intriguing. From the physics point of view, the fixed point FP3 corresponds to the pure HK model. Therefore, corresponding  critical dimensions should coincide with those in Eq.~\eqref{HKcrit}. Nevertheless, since $x^*=0$ at this point,  $\gamma^*_{\nu_\perp}$ also vanishes, so expression~\eqref{crete} gives us only the canonical part.

Usually we set $\beta_{\left\{g_i^*\right\}}=0$ in RG equation for any fixed points with the coordinates \mbox{$\left\{g_i^*\right\}=\left\{g^*,x^*,\alpha^*\right\}$}. In general, it is a correct step for all the values of $g^*$, $x^*$ and $\alpha^*$ except for zero and infinity. 
But here we need a more careful consideration. 
On the one hand, the variables $g$ and $x$ have well-defined finite limits for $g\to0$ and $x\to 0$ and, on the other hand, we are not interested in the limits $g\to\infty$ and $x\to\infty$. Therefore, we may put  $\beta_{g}=\beta_{x}=0$ for all the fixed points FP1~-- FP4. But we cannot make such a straightforward substitution for $\beta_\alpha$ at the fixed points with $\alpha^*=0$, i.e., at the fixed points which correspond to $u\to\infty$. Instead, 
we have to expand $\beta_\alpha$ up to the first nontrivial order and keep this term in the equation~(\ref{RGeqXX}).
Therefore, the expression analogous to~\eqref{G-scaling} 
at the fixed points FP3 and FP4 reads 
\begin{equation}
    \left({\cal D}_{k_\parallel}+{\cal D}_{k_\perp} +\Delta_\omega{\cal D}_\omega-
    \lambda^*{\cal D}_\alpha -d^k_G -\Delta_\omega d^\omega_G-\gamma^*_G\right)G^{R} = 0,
    \label{34-scaling}
\end{equation}
where $\lambda=\partial \beta_\alpha/\partial\alpha$ at $\alpha=0$, $\lambda^*$ denotes $\lambda(g^*,x^*)$, and $\Delta_\omega=2-\gamma_{\nu_\perp}^*$.

For the point FP3 where only the nonlinearity of the HK equation is relevant, the scaling equation takes the form
\begin{equation}
    \left({\cal D}_{k_\parallel}+{\cal D}_{k_\perp} +2{\cal D}_\omega-\frac{2}{3}\varepsilon\, {\cal D}_\alpha -d^k_G -2d^\omega_G-\gamma^*_G\right)G^{R} = 0.
    \label{3-scaling}
\end{equation}
The equation~\eqref{3-scaling} corresponds to scaling behavior with dilation of the momenta $k_\parallel$, ${\bf k}_{\bot}$, the frequency~$\omega$, and the ratio $\alpha$.
However, in this special case the model~\eqref{action_ve} coincides with the pure HK model~\eqref{action_pure} and acquires additional canonical symmetry which corresponds to independent dimensions $d^\parallel$ and~$d^\perp$, see~Eqs.~\eqref{ScInvX}.

The general solution of the set of two equations~(\ref{ScInv}),
RG equation~(\ref{RGeqXX}) taken at a fixed point, and
homogeneous counterpart of Eq.~\eqref{3-scaling} is an arbitrary function of three independent variables; for definiteness, we can choose them as 
\begin{equation}
    z_1=\frac{\omega}{\nu_\perp k_\perp^2}, \quad 
    z_2=\frac{k_\parallel}{k_\perp}, \quad \text{and} 
    \quad z_3=  \alpha \left(\frac{k_\perp}{\mu}\right) ^{2\varepsilon/3}.
    \label{sol1}
\end{equation}

Additional symmetry mentioned above means that the variables $z_1$, $z_2$, and $z_3$ should be dimensionless not only in respect to the Table~\ref{canon_dim_ve}, but in respect to the Table~\ref{canon_dim_pure}, too.
The variable~$z_1$ satisfies this requirement, variables $z_2$ and $z_3$ do not. 
Nevertheless, it is possible to construct the variable $z_0=z_2z_3^{-1/2}$ which has the needed canonical dimensions and is the second solution (along with $z_1$) of the homogeneous part of the equation~\eqref{HK-scalingXZ}:
\begin{equation}
    z_0= 
    \frac{k_\parallel }{k_\perp^{\Delta_\parallel} \alpha^{\wp}}\, \mu^{\varepsilon/3} 
    \quad \text{with} 
     \quad \wp=\frac{3}{2\varepsilon}\left(\Delta_\parallel-1\right),
    \label{sol2}
\end{equation}
where $\Delta_\parallel=1+\varepsilon/3$ is in agreement with expressions~\eqref{HKcrit}. This means, that the fixed point FP3 also
admits the scaling behavior, in which the coordinates $x_\parallel$ and ${\bf x}_{\bot}$ (i.e., momenta $k_\parallel$ and ${\bf k}_{\bot}$) 
are simultaneously rescaled with nontrivial $\Delta_\parallel\neq1$ while the IR irrelevant parameters (including $\alpha$) are kept fixed. Thus, the results derived for the pure HK model~\eqref{action_pure} are reproduced.

For the fixed point FP4 where both the nonlinearity of the HK equation and the turbulent advection are relevant, the 
equation of critical scaling~\eqref{34-scaling} reads
\begin{equation}
    \left({\cal D}_{k_\parallel}+{\cal D}_{k_\perp} +\Delta_\omega{\cal D}_\omega-
    \left(\frac{2}{3}\varepsilon-2\xi\right){\cal D}_\alpha -d^k_G -\Delta_\omega d^\omega_G-\gamma^*_G\right)G^{R} = 0,
    \label{4-scaling}
\end{equation}
where $\Delta_\omega=2-\xi$. Possible solutions of its homogeneous part are three functions $\hat{z}_1$, $\hat{z}_2$, and $\hat{z}_3$:
\begin{equation}
    \hat{z}_1= \frac{\omega}{\nu_\perp k_\perp^{2}}\left(\frac{k_\perp}{\mu}\right)^\xi, \quad 
    \hat{z}_2= \frac{k_\parallel}{k_\perp}, \quad
    \hat{z}_3= \alpha \left(\frac{k_\perp}{\mu}\right) ^{2\varepsilon/3-2\xi}.
    \label{FP4-an}
\end{equation}
The variables~\eqref{FP4-an} describe generalized critical scaling with rescaling of $\alpha$, 
i.e., the ratio of $\nu_\perp$ and~$\nu_\parallel$.
Any special case like~\eqref{sol2} can be obtained by combining $\hat{z}_1$, $\hat{z}_2$ and $\hat{z}_3$, but to perform it we need additional information about dependence of scaling function on these variables.
This situation resembles modified similarity hypothesis for systems which involve  different characteristic scales or different scaling laws; see~\cite{Stell,Stell2,Stell3,Stell4}.

\section{Conclusion. \label{sect:Con}}

We studied effects of isotropic turbulent advection described by Kazantsev--Kraichnan ``rapid-change'' ensemble~(\ref{white}) on the self-organized critical system modelled by the anisotropic Hwa--Kardar equation~(\ref{eq1})~--~(\ref{forceD}). 
The main reason to consider such problem is to understand how isotropy of turbulent flow and anisotropy of stochastic equation will interact with each other. 

We constructed a field theory equivalent to the problem under consideration and renormalized it. The renormalization constants were calculated in one-loop approximation~(\ref{zol}) but the critical dimensions were found exactly.

Four different universality classes of critical behavior were established: the regime of ordinary diffusion, the regime of the pure turbulent advection, the regime where only the nonlinearity of the Hwa--Kardar equation is relevant, and the regime where both the advection and the nonlinearity are relevant simultaneously. The most realistic values of the spatial dimension $d=2$ and the exponent $\xi=4/3$ correspond to the regime of the pure turbulent advection where the critical behavior is defined by the velocity ensemble. 

It was shown that  ``dimensional transmutation'' takes place in the regime where only the nonlinearity of the Hwa--Kardar equation is relevant. Precisely, the ratio $u$ of the two diffusivity coefficients $\nu_\parallel$ and $\nu_\perp$ acquires a nontrivial canonical dimension. 
Thus, the theory~\eqref{action_ve} obtains new canonical symmetry that corresponds to independent canonical dimensions $d^\parallel$ and $d^\perp$. 
As a result, this regime corresponds to the scaling behavior where the coordinates $x_\parallel$ and ${\bf x}_{\bot}$ are simultaneously rescaled with nontrivial exponent $\Delta_\parallel\neq1$ while the IR irrelevant parameters are kept fixed. This scaling behavior is in agreement with the one predicted by the pure Hwa--Kardar model~\eqref{action_pure}.

In the regime where both the advection and the nonlinearity are relevant, the scaling must involve rescaling of the ratio $u$. 
However, additional information about scaling function is required to calculate the critical exponents.
This result brings to mind modified similarity hypotheses 
(weak scaling in the spirit of Stell~\cite{Stell,Stell2,Stell3}
and generalized scaling in the spirit of Fisher~\cite{Stell4}) for systems with different characteristic scales or different scaling laws.

\funding{
The reported study was funded by RFBR, project number~20-32-70139.
The work by N.~V.~Antonov and P.~I.~Kakin was also supported by the Foundation for the Advancement of Theoretical Physics and Mathematics~``BASIS.''}


\pagebreak

\abbreviations{The following abbreviations are used in this manuscript:\\

\noindent 
\begin{tabular}{@{}ll}
HK & Hwa--Kardar \\
IR & infrared \\
MS & mimimal subtraction \\
NS & Navier--Stokes \\
RG & renormalization group \\
SOC & self-organized criticality \\
UV & ultraviolet \\
\end{tabular}}

\reftitle{References}

\end{document}